\documentclass{article}
\usepackage{float} 
\usepackage[sumlimits]{amsmath}
\usepackage{multicol}

\usepackage{formatka}

\input{epsf}
\newfloat{Fig.}{H}{w}

\newcommand{\be}{\begin{eqnarray}}
\newcommand{\ee}{\end{eqnarray}}

\begin{document}

\title{Combining Tight-binding and Molecular dynamics Methods to Model 
       the~Behaviour of~Metals \\in~the~Plastic Regime} 

\author{Maciej Bobrowski$^{1,2}$, Jacek Dziedzic$^{1}$, Jaros\l{}aw Rybicki$^{1,2}$}

\address{$^{1}$Faculty of~Technical Physics and Applied Mathematics, \\
               Gda\'{n}sk University of~Technology, \\
          Narutowicza~11/12, 80-952 Gdansk, Poland \\
}

\email{Phone: 0048 58 347 28 34, e-mail: ryba@pg.gda.pl
}

\address{$^{2}$TASK Computer Centre, \\
               Gda\'{n}sk University of~Technology, \\
          Narutowicza~11/12, 80-952 Gdansk, Poland \\
}

\noindent
{\bf Abstract}\\
Ultra-precision machining of metals, the breaking of nanowires under tensile stress 
and fracture of nanoscale materials are examples of technologically important
processes which are both extremely difficult and costly to~investigate 
experimentally. We describe a multiscale method for the simulation of such systems in which 
the~energetically active region is modelled using a robust tight-binding scheme developed
at the Naval Research Laboratory (NRL-TB) and the rest of the system is treated
with molecular dynamics.
We present a computer code implementing the method, geared towards non-equilibrium,
cross-scaled tight-binding and molecular dynamics simulations.
Apart from the presentation of the method and implementation, we discuss preliminary physical 
results obtained and discuss their validity.

\begin{multicols}{2}

\section{Introduction}
Ultraprecision machining, or~UPM, is~a~process in~which a~hard (usually diamond or~CBN) 
tool machines away a~layer of~workmaterial as~thin~as~a~few nanometers. The~machined 
materials include soft metals, such as~aluminum or~copper, glasses and semiconductors.
A~prime example of~a~technologically important application of~UPM is~the~finishing 
of~aluminum platters for~magnetic hard disk drives \cite{Wear1}, \cite{Wear2}. 
Experimental analysis~of~UPM is~both extremely difficult and costly to investigate:
the~need to damp vibrations, assuring adequate vacuum and the~necessity to use AFM
or~STM to visualize the~results being the~most important obstacles. 

High costs of~experimental setups have led researchers to explore the~possibilities
of~computer simulation of~ultraprecision machining processes. Molecular dynamics
(MD) is~the~traditional method of~choice because of~its relative simplicity
and good scalability. The~increase of~available computational power offers today
the~opportunity to perform MD simulations of~UPM using $10^4$-$10^5$ atoms with 
timescales in~the~order of~a~few nanoseconds \cite{Wear1}, \cite{Wear2}, \cite{Monika}.

Such simulations are performed in the non-equi\-lib\-rium (NEMD) regime, with the
tool atoms usually having their velocity reset to a~constant value at each step. The
workmaterial needs artificial "clamping" to~ensure it remains in~place and
does not translate upon contact with the~tool. Periodic boundary conditions
may \cite{Monika} or~may not \cite{Wear1} be used along the~direction 
perpendicular to that of~the~machining. The~potentials used are either pairwise,
such as~Morse \mbox{potential} \cite{Wear2} or~many-body, such as~the~Sutton-Chen
potential \cite{Monika}. Since these are parametrized to reproduce the~behaviour
of~atoms close to their equilibrium positions, they are rather unreliable
in~situations where bonds are ruptured and such is~the~case in
the~part of~the~system where the~tool tip enters the~material. In~fact it was
shown that for~brittle fracture of~silicon the~widely used Stillinger-Weber
MD potential gives incorrect predictions, as~compared to DFT-based methods \cite{LOTF_ref1}.

Apart from UPM there are other interesting systems and processes which
also do not lend themselves easily to experimental investigation and
contain~regions where bond-breaking takes place, which renders MD
simulations thereof~unreliable. Among these are: metallic nanowires breaking
under stress, nanoindentation, nano-scratching, fracture of~nanoscale
materials. If the~system is~small (less than $10^3$ atoms) quantum-based 
methods, such as~DFT or~tight-binding (TB) which deal with the
electronic structure explicitly may be used. For~larger systems the
computational cost of~these methods quickly becomes prohibitive.
There is, however, a great need for methods that would allow for
the treatment of such systems, thus allowing to simulate more complex materials, auxetics being
one of the greatest challenges.

One solution of~this~problem is~offered by~the~so-called cross-scaling
methods, in~which a~part of~the~system that needs quantum-based treatment
is~identified and treated accordingly, while the~remaining part of~the
system is~simulated using the~MD method. This~approach attempts to benefit from
the~advantages of~both methods, however it suffers from its own problems,
such as~difficulty in~determining the~position of~the~region and dealing
with boundaries between the~two partitions of~the~system. Handshaking
the~two methodologies is the~most enduring task. 

The~paper is~organized as~follows. In~section~2 we~describe the~formulation
of~the~tight-binding method and the~molecular dynamics method, later
focusing on~bridging of~the~two methods. Our computer code implementing
the~cross-scaling is~then characterized, along with the~simulation procedure.
In~section~3 we~present results of~preliminary tests used to~validate the~method.
Section~4 contains conclusions and final remarks. 

\section{Computational method}

\subsection{The~NRL-TB method}
Tight-binding is~the~only quantum-based method that allows for~the~treatment
of~systems comprising several hundred atoms on a~typical workstation. While
relatively simple compared to DFT-based methods, it still captures the
relevant electronic efects, at least in~a~qualitative manner. The~method
reduces the~Schr\"{o}dinger equation to a~generalized eigenvalue problem
by~expanding one-electron wavefunctions as~linear combinations of~atomic orbitals.
Following Slater and Koster \cite{SlaterKoster} matrix elements of~the
hamiltonian are separated into an angular-dependent part and distance-dependent
two-center integrals, which are then parametrized. Brute-force matrix diagonalization
which is~involved in~solving the~eigenvalue problem scales as~$O(N^3)$ and is
usually the~bottleneck of~the~calculations.	 $O(N)$-scaling methods have also
been developed, but since they rely on bond locality, they are less transferable 
and perform poorly for~metals \cite{Colombo}. In~the~tight-binding formalism the~energy 
of~the~system is~written as:
\be
E_{TB} = 2 \sum_{n}^{N_{occ}} \varepsilon_n + F\left[n(r)\right],
\ee
where the~first term, involving summation of~the~electronic eigenvalues $\{\varepsilon_{n}\}$ over
all occupied levels is~the~band structure energy. The~second term is~an~unknown functional of~the~charge
density $n(r)$ and accounts for~the~remaining density-functional energy (coulombic core-core repulsion,
canceling out the~double counting of~electron-electron interactions, etc.) This~second term is~usually
approximated by~a~parametrized repulsive pair potential, which necessitates using extra~parameters 
apart from the~ones describing the~two-center integrals.

In~the Naval Research Laboratory tight-binding (NRL-TB) total-energy method \cite{P1}-\cite{P15}, developed by~Mehl and Papaconstantopoulos
the~second term is~rid of~by~clever shifting of~the~underlying DF Kohn-Sham potential. This~narrows
down the~parameter space at the~price of~introducing a~dependence of~the~on-site matrix elements
on the~local environment. It can be shown \cite{P1}-\cite{P3} that such approach is~well-justified and is~a~generalization
of~the~pair potential typically used for~$F\left[n(r)\right]$. The~\mbox{NRL-TB} method offers transferability
superior~to other TB variants and parameter sets for~many $d$-metals are readily available. 
These advantages have made it~our method of~choice for~the~quantum-based region of~the~system, 
despite the~fact that the~computer code implementing the~method is~no longer given away by~the~authors.

\subsection{The~molecular dynamics method}
In~the~MD method a~system of~classical particles interacting with one another via~an empirical
potential is~followed in~discrete timesteps. Analytical differentiation of~the~potential yields 
forces acting on~the~particles and numerical integration of~Newtonian equations of~motion gives
positions of~the~particles in~subsequent timesteps. Aided with the~link-cell method, it scales
as~$O(N)$ and simulations of~hundreds of~thousands of~particles for~timescales of~a~few
nanoseconds are feasible on a~typical workstation. A~usual timestep is~in~the~order of~1 fs,
we~have chosen $\Delta t=2.5$~fs. We~have decided to choose a~fourth-order Gear predictor-corrector
algorithm as~the~numerical integrator. 

\subsection{nanoTB code}
After careful investigation of~available non-com\-mer\-cial MD codes we have decided that none of
them met our needs for~simulating ultraprecision machining or~other systems involving
plastically deforming metals. However a~parallel code suitable for~performing NEMD simulations was~being
developed in~our group as~part of~research on carbon nanotubes \cite{Michal_PTSK}. This~code
was~easily adapted to our task of~interest by~the~introduction of~the~Sutton-Chen many-body potential. 
The~nanoMD program implements the~molecular dynamics method and is~geared towards simulations
involving external forces, therefore making it easy to fix atoms in~position, apply constant
translatory or~rotational forces, etc. A~selection of~potentials, including Morse, Sutton-Chen,
Brenner, Finnis-Sinclair, harmonic and anharmonic is~available. Nos\'{e}, Nos\'{e}-Hoover and gaussian
thermostats have also been implemented. It has~been used with success for~the~MD simulation 
of~UPM \cite{Monika}.

To perform cross-scaling simulations we have implemented the~NRL-TB method from scratch,
incorporating it into our code. The~program, now called nanoTB, accepts the~TB parameter files available from the~NRL website. 
Two modes of~operation are possible -- in~the~first a~TBMD simulation is~performed, that
is~the~whole system is~treated with the~TB method, and the~quantum-based forces it supplies
drive the~MD simulation. The~second mode allows for~a~cross-scaling TB+MD simulation -- regions
of~the~system are designated for~which a~TB calculation takes place, in~these regions the~quantum-based
forces replace the~forces obtained from MD, while the~rest of~the~system is~treated with MD. Forces
are obtained from the~TB hamiltonian matrix element derivatives by~means of~Hellmann-Feynman
theorem. Each region may take a~spherical, cylindrical or~cuboid shape and be either fixed in
size or~dynamically adjust to accommodate a~fixed number of~atoms. For~the~cuboid and cylindrical
region shapes periodic or~fixed boundary may be used. Charge self-consistency may be achieved 
by~means of~Hubbard-U method or~be neglected. Fermi broadening of~energy levels is~used. The~regions
may be fixed in~place, move with uniform velocity (eg. along with the~tool tip) or~be made to 
follow energy peaks in~the~system (useful for~tracing defects). 

\subsection{Handshaking TB and MD}
Devising a~physically sound method for~embedding the~\textit{ab-initio} 
region within~an MD system is~of~utmost concern for~cross-scaling simulations.
Spurious effects resulting from the~cessation of~bonds at the~TB/MD interface as
the~TB region is~isolated from the~rest of~the~system have plagued cross-scaling simulations 
from their inception \cite{Kaxiras}, \cite{skale_chemikow1}, \cite{skale_chemikow2}. 
The~approach usually taken to account for~the~unsaturated valencies is~to surround 
the~quantum region with virtual atoms, usually monovalent, which serve to
terminate the~dangling bonds \cite{Kaxiras}, \cite{skale_chemikow2}. Since they introduce extra~atoms 
into the~calculation, these link-atom methods (LAM) increase the~computational load and
require a~parametrization of~the~method not only for~the~species of~interest but also
for~the~interactions between it~and the~link-atoms. Even more important is~the~fact
that they rely on bond locality which prohibits their use for~metallic systems. In~fact
it seems that no handshaking method that could deal with delocalized bonding has~ever
been proposed.

Our first attempt to approach this~problem was~as~follows. We have conceded to the~fact
that unterminated valencies will be present in~the~TB region but have instead concentrated
on reducing their influence on the~atomic trajectories. Assuming that the~greatest
disturbances in~the~TB-generated forces would be present in~the~outer part of~the~quantum
region, we decide to give more credence to the~MD forces for~the~atoms close to the
region perimeter. This~was~achieved by~taking for~the~force acting on atom~$i$ within~the
region a~weighted average of~the~TB force and the~MD force, the~weight being the~relative
distance $d$ of~the~atom from the~region centre:
\be
\vec{F}_i = d \cdot \vec{F}^{MD}_i + (1-d) \vec{F}^{TB}_i.
\label{eq:linear}
\ee
This~procedure was~meant to result in~a~gradual switching from the~TB forces to the~MD forces
as~the~former became more unreliable. 

Unfortunately this~simple approach has~led to serious unphysicalities in~the~test cases
we have studied. Upon centering a~spherical TB region of~14~\AA{} in~diameter 
on~a~self-interstitial defect in~Cu, the~quantum-based forces acting on the~atoms close to 
the~perimeter of~the~region were two to three orders of~magnitude greater than 
the~corresponding MD forces. This~was~due to the~fact that these atoms were only very 
slightly disturbed from their lattice positions, yet they were heavily influenced 
by~the~isolation of~the~TB region from the~rest of~the~system. As~a~consequence, even
with $d=0.95$ the~force computed from (\ref{eq:linear}) was~overestimated by~an
order of~magnitude or~so for~all atoms close to the~region boundary. A~similar
scenario occured for~a~cylindrical region centered on~one of~the~edges of~a~cuboid 
of~1638 copper atoms, mimicking a~workmaterial to~be machined. The~strong TB forces
resulting from the~isolation of~the~region easily outweighed the~relatively small
MD forces acting on these same atoms. As~the~hamiltonian resulting from (\ref{eq:linear})
is~not conservative, this~resulted in~constant increase of~system energy, effectively
melting the~surrounding material.

Thus it became obvious that a~more sophisticated method of~combining the~TB and MD forces
was~needed. We have settled for~a~method similar in~concept to the~one described, but using a
more aggressive mixing in~of~the~MD forces. We have decided to cut-off the~influence of
the~TB forces using a~smooth function similar to the~cutoff function used in~the~NRL-TB method \cite{P4}.
Instead of~(\ref{eq:linear}) we use
\be
\vec{F}_i = w(d) \vec{F}^{MD}_i + w(1-d) \vec{F}^{TB}_i,
\label{eq:nonlinear}
\ee
where
\be
w(d)=\dfrac{1}{1+\exp{(-\frac{d}{l}}+5)},
\ee
with $l=0.06$. This~leads to a~non-linear weighing of~the~TB and MD
forces. 
\begin{Fig.}
\epsfxsize=7cm
\epsfbox{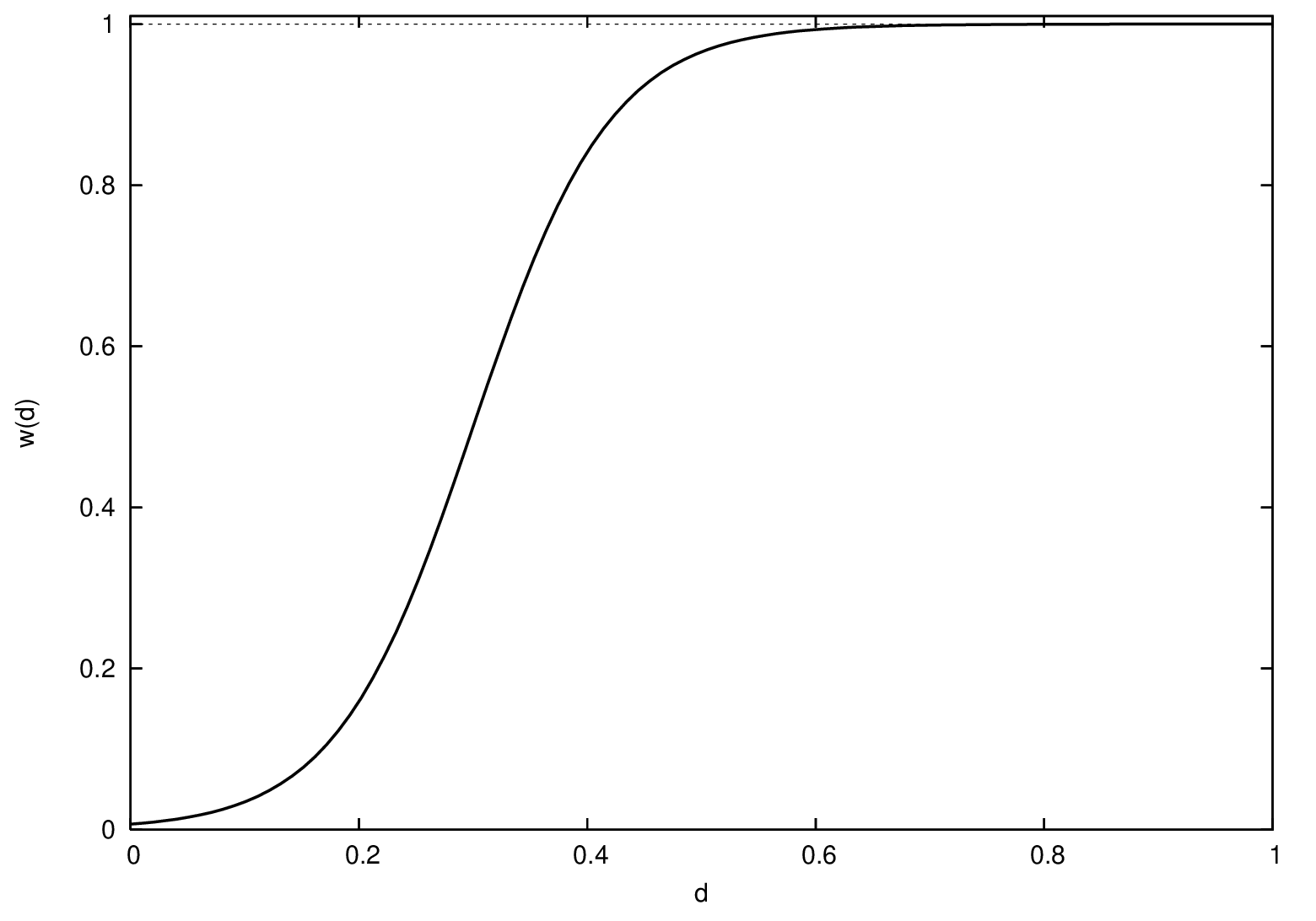}
\caption{{The~shape of~the~weighing function $w(d)$. \newline Note $w(0) \approx 0$, $w(1) \approx 1$.}}
\end{Fig.}
This~improved embedding has~proven more successful against our test cases. As~an example
consider a~cylindrical quantum region of~diameter 13.25 \AA{} placed on~a~surface of~a~cuboid, with
the~cylinder axis~parallel to the~cuboid edge. This~system consisting of~4305 total atoms and 120
TB atoms was~simulated for~8~ps and results of~the~two mixing methods were compared.
\begin{Fig.}[H]
\epsfxsize=6.5cm
\epsfbox{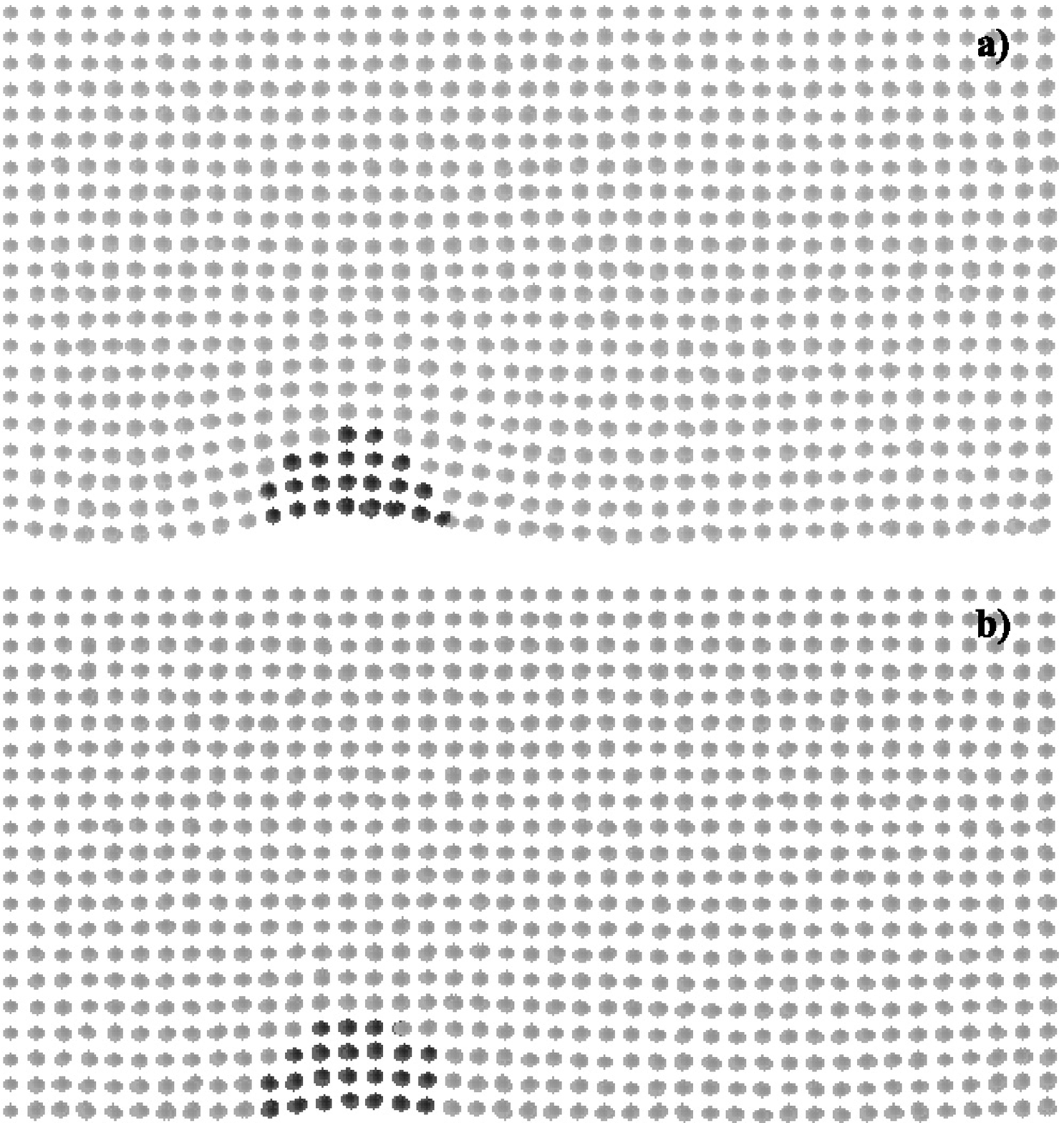}
\caption{Artifacts introduced by~unterminated valencies of~the~TB region using the~(a) linear; (b) nonlinear 
mixing method. A~snapshot of~the~system configuration at $t=0.85$ ps is~shown, with the~TB atoms drawn
in~darker colour.}
\end{Fig.}
The~artifacts introduced by~the~nonlinear method were markedly smaller, although still visibly present.
This, naturally, comes at the~cost of~having to extend the~region in~practical applications
as~the~atoms further than $d=0.5$ from the~region centre are effectively driven 
by~the~MD method, however their inclusion in~the~TB calculation serves to create a~proper local quantum
environment for~the~atoms closer to the~region centre.

\subsection{Simulation procedure}
As~a~preliminary test of~our code and the~nonlinear mixing method, we have performed a
cross-scaled simulation of~nanoscratching of~a~copper workmaterial with a~copper tool.
The~workmaterial was~a~cuboid of~20x10x5 fcc unit cells, for~a~total of~4305 atoms. The~tool 
consisted of~a~cuboid of~8x8x5 fcc unit cells (1445 atoms), rotated by~45~degrees along the~$z$ axis~and 
moving along the~$[0\overline{1}0]$ direction. Periodic boundary conditions were used
along the~$z$ (in-plane) direction. Atoms on the~bottom surface of~the~workmaterial
were fixed to prevent it from translating upon contact with the~tool. For~simplicity
the~tool was~assumed to be infinitely hard (all forces acting upon it were ignored.)
the~tool moved with a~uniform velocity of~10 $\textrm{m}/\textrm{s}$, which is
considerably more realistic than in~other simulations of~UPM \cite{Wear1},
\cite{Wear2}. A~pure MD simulation using the~Sutton-Chen potential was~performed for
comparison. The~quantum region was~cylindrical with periodic boundary conditions along the~$z$ axis
and was~centered on the~tool tip, moving with the~same velocity as~the~tool. It~was taken to 
be only 8 \AA{} in~diameter for~efficiency reasons, in~realistic simulations the~region should have
a~radius of~at least 6.62 \AA{}, since this~is~the~cutoff distance for~Cu in~the~NRL-TB method. 
Since the~region was~fixed in~size, the~number of~atoms contained within~it varied with time,
with an average of~70 atoms. The~simulations were performed at 300 K with a~Nos\'{e}-Hoover thermostat
having a~time constant of~250 fs. A~timestep of~2.5 fs was~used. Initially the~tool tip was
at a~distance of~4.75 \AA{} from the~workmaterial, giving it ample time to equilibrate before
contact was~made.

\section{Results and discussion}
We consider the~presented simulation to be merely a~preliminary test, therefore we look at the~obtained
results only qualitatively. The~differences between the~cross-scaled and the~reference, pure MD simulation
are rather subtle, but noticeable. For~both simulations we observe the~jump-to-contact (JC) phenomenon
when the~distance between the~tool tip and the~workmaterial is~about 4 \AA. This~is~in~agreement with
pure MD simulations described in~\cite{Wear1}, \cite{Wear1_refLandman1}, \cite{Wear1_refLandman2}.
Since the~tool is~infinitely rigid, it is~the~workmaterial that deforms slightly to contact the~tool.
At this~point the~cross-scaling and the~reference simulation are expected to perform identically,
because as~the~TB region is~centered on the~tool tip, at this~distance there are only 15 workmaterial
atoms within~the~region and these are driven almost exclusively by~MD. 
\begin{Fig.}[H]
\begin{center}
\epsfxsize=6.5cm
\epsfbox{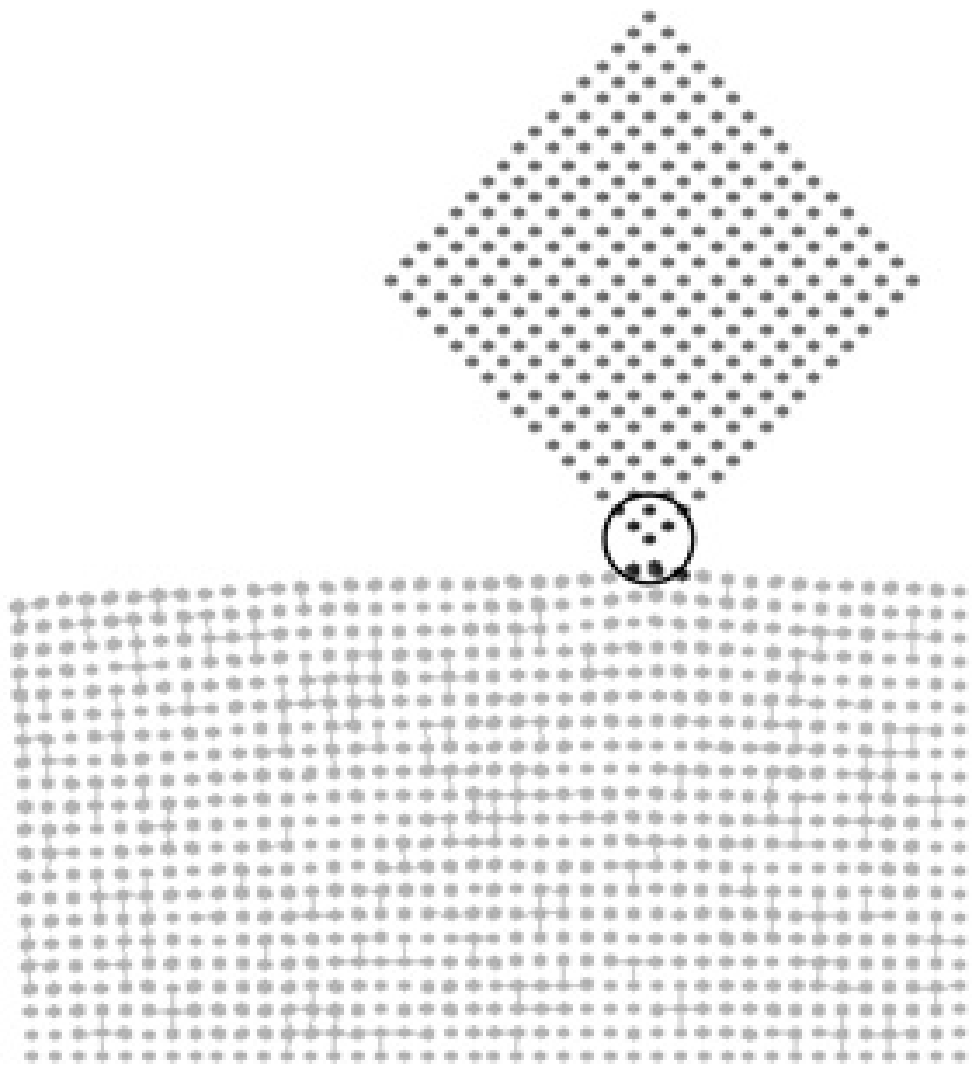}
\caption{The~jump-to-contact phenomenon. The~TB region is~marked with a~circle with the~TB atoms drawn
in~a~darker colour. Note how this~effectively reduces to an MD simulation as~the~workmaterial atoms
have only begun to enter the~quantum region.}
\end{center}
\end{Fig.}
\begin{Fig.}
\begin{center}
\epsfxsize=7.5cm
\epsfbox{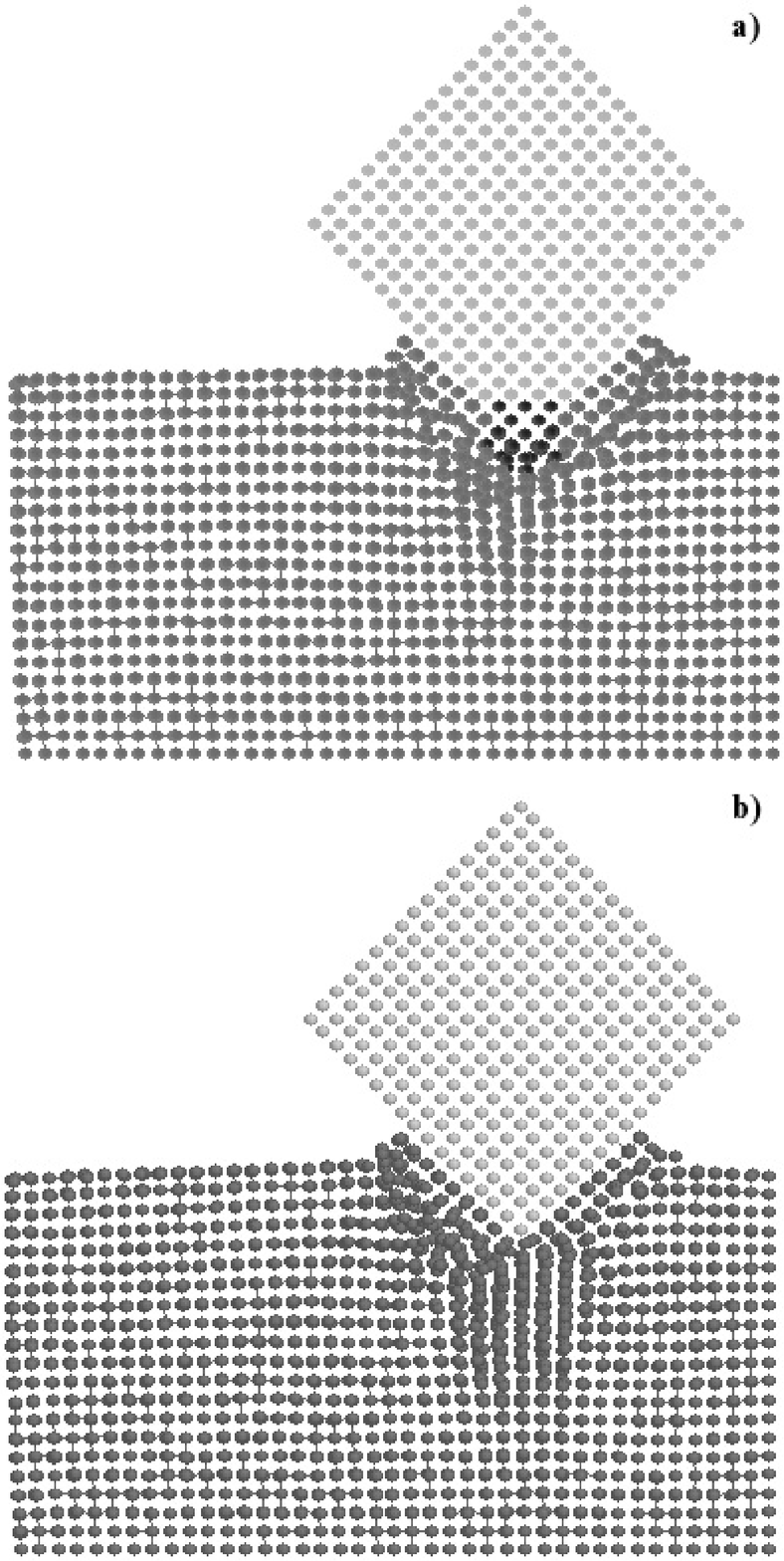}
\caption{A snapshot of the configuration at $t=100$ ps, after the tool has penetrated
the workmaterial 5.25 \AA{} deep. a) cross-scaling TB+MD simulation (TB atoms shown in darker
colour); b) pure MD simulation. }
\end{center}
\end{Fig.}
\vspace{-0.35cm}
Until the~distance between the~tool tip and the~workmaterial reaches about 1.5 \AA{},
the~values for the~normal force acting on~the~tool are practically identical in~the~two
simulations and extremely small -- the~tool is~attracted by~the~workmaterial with 
a~force lower than 0.1 nN. This force fluctuates quasi-periodically as~the~workmaterial "wiggles" 
under the~tool in~response to~the~attraction. As~the~tool tip gets closer than 1.5~\AA{} to~the~surface,
the~force becomes repelling and then rises steadily after contact as~the~tool penetrates the~workmaterial. 
The~forces achieve their highest value of~about 0.8 nN at~the~penetration depth of~8~\AA{} (where
the~simulation terminates).
During the~course of~the~simulation, as~more atoms are treated with TB, the~differences between the~simulations
become noticeable.
The~normal forces observed in~the~cross-scaling simulation are seen to~be~10-30\% greater than
the~corresponding pure MD forces.

Taking a~closer look at the~configuration of~the~system after the~tool has~entered the
workmaterial 5.25 \AA{} deep, we note slight differences between the~cross-scaling and
reference simulation. A~comparison of~the~figures below reveals that the~cross-scaled
simulation shows a~smaller disturbance of~workmaterial several layers under the~tool tip.
Slight differences in~the~pressure maps (Fig. 5) and the~fact that the~normal force
experienced by~the~tool is~greater for~the~cross-scaling simulation lead to a~supposition
that the~material is~in~fact slightly harder than pure MD simulations would predict. 

\begin{Fig.}
\begin{center}
\epsfxsize=6.4cm
\epsfbox{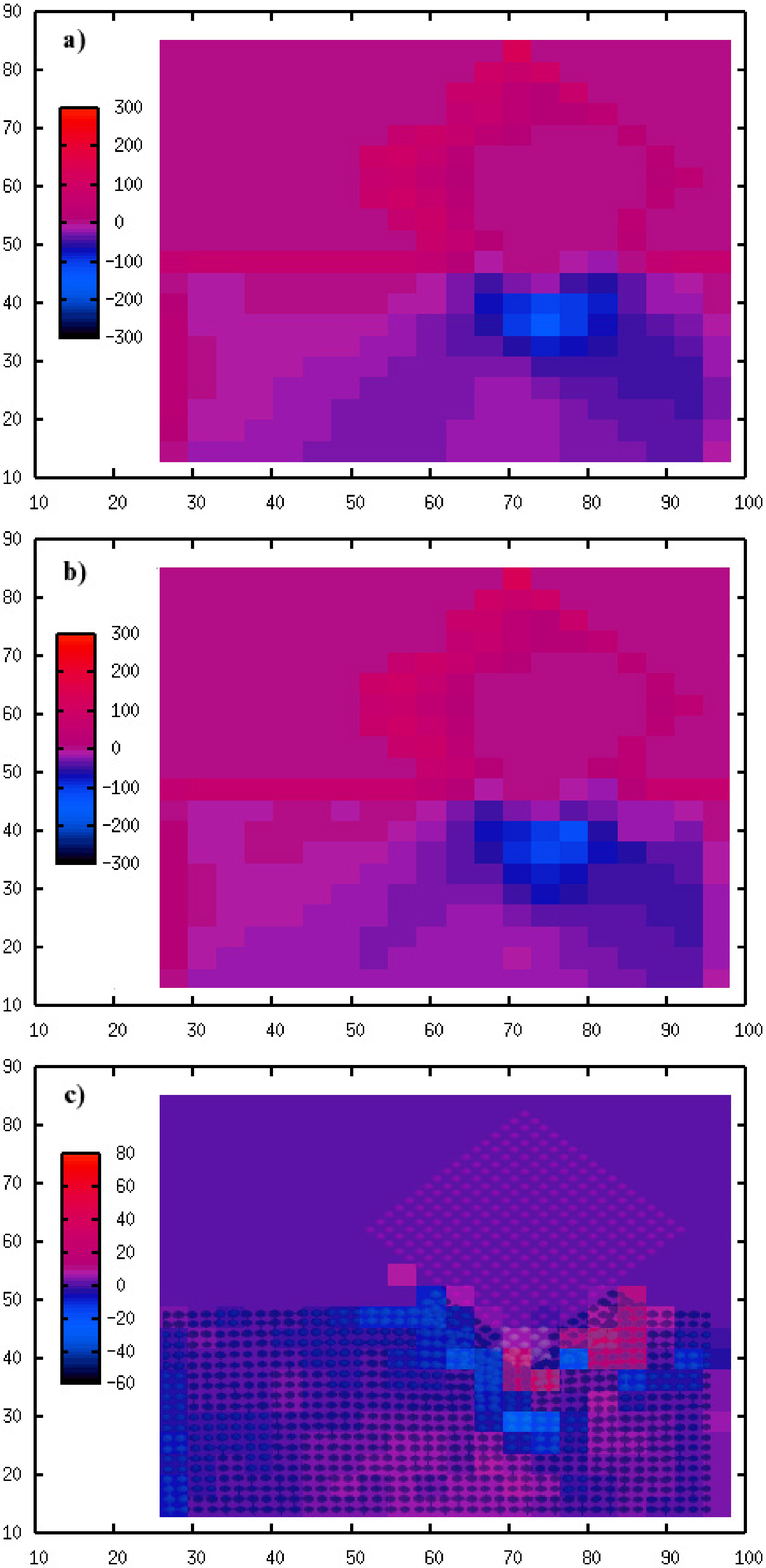}
\caption{Pressure map (arbitrary units) for the snapshot in Fig. 4. a) cross-scaling TB+MD simulation; b) pure MD 
simulation; c) the difference between the~two (image of system configuration overlapped for clarity). }
\end{center}
\end{Fig.}
\section{Conclusion}
We~have developed a~computer code that allows for cross-scaling simulations using the~methods 
of~NRL total energy tight-binding and molecular dynamics in~combination. The~program allows for
the~embedding in~an~NEMD simulation a~number of~quantum-based regions that are dynamic in~size 
and position and may take various shapes. We~have considered two methods that would tentatively
deal with the~problem of~unterminated valencies in~metallic systems on~the~TB/MD boundary,
showing the~first one to~fail and the~second to~be~moderately successful. The~nonlinear
mixing method minimizes the~surface effects on~region boundaries at~the~price of~having 
to~extend the~region, which increases the~computational load. Also the~hamiltonian resulting
from the~application of~the~method is~not conservative. We~have tested the~code and 
the~nonlinear cross-scaling method on~several simple systems, finding the~code to~work and 
the~method to~behave better than the~linear one we~have proposed earlier. For the~simulation 
of~nanoscratching, our cross-scaling method gave similar, but not identical results 
to~the~pure MD simulation, we~bear in~mind, however, that the~region size used for the~test
was too small for practical applications. We~envisage adapting the~recent learn-on-the-fly
method \cite{LOTF} to~the~Sutton-Chen potential in~order to~attack the~problem of~embedding
quantum-based calculations within an~MD~framework from a~different perspective. 

\vskip 0.5cm
\noindent
{\bf Acknowledgements}

The~simulations have been performed at the~TASK Computer Centre (Gdansk, Poland).
The~work has~been sponsored by~KBN, under grant number 7~T11F~013~21.


\end{multicols}


\noindent

\begin{thebibliography}{99}

\bibitem{Wear1} R.~Komanduri, N.~Chandrasekaran and L.~M.~Raff, \textit{MD~simulation of~indentation and scratching of single crystal aluminum}, Wear \textbf{240}, 113 (2000).
\bibitem{Wear2} R.~Komanduri, N.~Chandrasekaran and L.~M.~Raff, \textit{M.D.~Simulation of~nanometric cutting of~single crystal aluminum -- effect of~crystal orientation and direction of~cutting}, Wear \textbf{242}, 60 (2000).
\bibitem{Monika} M.~Bia\l{}osk\'{o}rski, M.~Rychcik-Leyk, J.~Rybicki, G.~Bergma\'{n}ski, \textit{Molecular dynamics simulations of~the~ultraprecision cutting of~metals}, Proc. of~the~9-th Workshop of~PTSK Koszalin-Osieki \textbf{9} (2002).
\bibitem{Michal_PTSK} M.~Bia\l{}osk\'{o}rski and J.~Rybicki, \textit{Mechanical properties of~the~carbon nanotubes: simulation program and exemplary results}, Proc. of~the~8-th Workshop of~PTSK Gda\'{n}sk-Sobieszewo \textbf{8} (2001).
\bibitem{Kaxiras} J.~Q.~Broughton, F.~F.~Abraham, N.~Bernstein and E.~Kaxiras, \textit{Concurrent coupling of~length scales: methodology and application}, Phys. Rev. B \textbf{60}, 2391-2403 (1999).
\bibitem{LOTF_ref1} N.~Bernstein and D.~W.~Hess, \textit{Lattice trapping barriers to~brittle fracture}, Phys. Rev. Lett. \textbf{91}, 025501 (2003).
\bibitem{SlaterKoster} P.~Slater and G.~Koster, \textit{Simplified LCAO Method for the~Periodic Potential Problem}, Phys. Rev. \textbf{94}, 1498 (1954). 
\bibitem{Colombo} L.~Colombo, \textit{Tight-binding molecular dynamics}, Annual Reviews of Computational Physics IV, 1996. 
\bibitem{P1} M.~Mehl and D.~Papaconstantopoulos, \textit{Tight-binding parametrization of~first-principles results}, Topics in Computational Materials Science, C. Y. Fong, ed. (World Scientific, Singapore, 1998) Ch. V, pp. 169-213.
\bibitem{P2} D.~Papaconstantopoulos and M.~Mehl, \textit{The~Slater-Koster Tight-Binding Method: a~Computationally Efficient and Accurate Approach}, unpublished.
\bibitem{P3} D.~Papaconstantopoulos and M.~Mehl, \textit{Applications of~a~New Tight-Binding Total Energy Method}, Proc. of the International Symposium on Novel Materials, Bhubaneswar, India, March 3-7, 1997, edited by B.K. Rao (1998), pp. 393-403.
\bibitem{P4} R.~Cohen, M.~Mehl and D.~Papaconstantopoulos, \textit{Tight-binding total-energy method for~transition and noble metals}, Phys. Rev. B \textbf{50}, 14694-14697, 1994. 
\bibitem{P5} M.~Mehl and D.~Papaconstantopoulos, \textit{Application of~a~new Tight-Binding method for~transition metals: Manganese}, Europhys. Lett. 31 \textbf{537}, (1995).
\bibitem{P6} M.~Mehl and D.~Papaconstantopoulos, \textit{Applications of~a~tight-binding total-energy method for~transition and noble netals: Elastic constants, vacancies and surfaces of~monatomic metals}, Phys. Rev. B \textbf{54}, 4519 (1996). 
\bibitem{P7} S.~Yang, M.~Mehl and D.~Papaconstantopoulos, \textit{Application of~a~tight-binding total-energy method for~Al, Ga, and In}, Phys. Rev. B, \textbf{57} R2013 (1998). 
\bibitem{P8} M.~Mehl, D.~Papaconstantopoulos, N.~Kioussis and M.~Herbranson, \textit{Tight-binding study of~stacking fault energies and the~Rice criterion of~ductility in~the~fcc metals}, Phys. Rev. B \textbf{61}, 4894 (2000). 
\bibitem{P9} N.~Bernstein, M.~Mehl, D.~Papaconstantopoulos, N.~Papanicolaou, M.~Bazant and E.~Kaxiras, \textit{Energetic, vibrational, and electronic properties of~silicon using a~nonorthogonal tight-binding model}, Phys. Rev. B \textbf{62}, 4477 (2000). 
\bibitem{P10} F.~Kirchhoff, M.~Mehl, N.~Papanicolaou, D.~Papaconstantopoulos and F.~Khan, \textit{Dynamical properties of~Au from tight-binding molecular-dynamics simulations}, Phys. Rev. B \textbf{63}, 195101 (2001). 
\bibitem{P11} G.~Wang, D.~Papaconstantopoulos and E.~Blaisten-Barojas, \textit{Pressure induced transitions in~calcium: a~tight-binding approach}, J. Phys. Chem. Sol. \textbf{64}, 185 (2002). 
\bibitem{P12} D.~Papaconstantopoulos, M.~Mehl, S.~Erwin and M.~Pederson, \textit{Tight-binding hamiltonians for~carbon and silicon}, Tight-Binding Approach to Computational Materials Science, P.~Turchi, A.~Gonis, L.~Colombo, ed., MRS Proceedings \textbf{491}. 
\bibitem{P13} Y.~Mishin, M.~Mehl, D.~Papaconstantopoulos, A.~Voter and J.~Kress, \textit{Structural stability and lattice defects in~copper: \emph{Ab initio}, tight-binding, and embedded-atom calculations}, Phys. Rev. B \textbf{63}, 224106 (2001). 
\bibitem{P14} D.~Papaconstantopoulos, M.~Lach-hab and M.~Mehl, \textit{Tight-binding Hamiltonians for~realistic electronic structure calculations}, Physica~B \textbf{296}, 129 (2001). 
\bibitem{P15} N.~Bernstein, M.~Mehl and D.~Papaconstantopoulos, \textit{Nonorthogonal tight-binding model for~germanium}, Phys. Rev. B \textbf{66}, 075212 (2002). 
\bibitem{Wear1_refLandman1} U.~Landman and W.~D.~Luedtke, \textit{Nanomechanics and dynamics of~tip-substrate interactions}, J. Vac. Sci. Technol \textbf{9}, 414 (1991).
\bibitem{Wear1_refLandman2} U.~Landman and W.~D.~Luedtke, \textit{Atomistic dynamics of~interfacial processes: films, junctions and nanostructures}, Appl. Surf. Sci. \textbf{92}, 237 (1996).
\bibitem{skale_chemikow1} X.~P.~Long, J.~B.~Nicholas, M.~F.~Guest and R.~L.~Ornstein, \textit{A~combined density functional theory/molecular mechanics formalism and its application to small water clusters}, J. Mol. Struct. \textbf{412}, 121 (1997).
\bibitem{skale_chemikow2} M.~Eichinger, P.~Tavan, J.~Hutter and M.~Parrinello, \textit{A~hybrid method for solutes in~complex solvents: Density functional theory combined with empirical force fields}, J. Chem. Phys. \textbf{110}, 10452 (1999).
\bibitem{LOTF} G.~Cs\`{a}nyi, T.~Albaret, M.~C.~Payne and A.~De Vita, \textit{"Learn on the Fly": A~Hybrid Classical and Quantum-Mechanical Molecular Dynamics Simulation}, Phys. Rev. Lett., \textbf{93}, 175503 (2004).
\end{thebibliography}
\end{document}